\documentclass{article}
\usepackage{epsfig,graphics,amssymb,amsmath,amsthm,subeqnarray,color,multirow,}
\usepackage{JMR}

\newcommand{\beq}{\begin{equation}}
\newcommand{\eeq}{\end{equation}}

\def \bu{\mathbf{u}}

\def \Re{\text{Re}}
\def \Pe{\text{Pe}}
\def \Cx{\text{Cx}}

\def \d{\text{d}}

\def \bF{\mathbf{F}}
\def \cF{\mathcal{F}}

\def \bI{\mathbf{I}}

\def \bx{\mathbf{x}}
\def \bk{\mathbf{k}}
\def \bu{\mathbf{u}}

\def\bbf{\mathbf{f}}

\def \bI{\mathbf{I}}

\def \bx{\mathbf{x}}

\def \bxh{\hat{\boldsymbol{x}}}
\def \byh{\hat{\boldsymbol{y}}}
\def \bzh{\hat{\boldsymbol{z}}}

\def \log{\, \mathrm{log} \,}

\def \bpsi{\boldsymbol{\psi}}

\def \tilde{\widetilde}
\def \d{\, \text{d}}
\def \e{\epsilon}
\def \tot{\text{tot}}
\def \R{\mathbb{R}}
\def \ksp{\bk-\text{space}}
\def \Pavg{\overline{P}}
\def \units{\, \mathrm }

\def \llangle{\left \langle}
\def \rrangle{\right \rangle}

\renewcommand{\bx}{\ensuremath {\boldsymbol {x}}}
\renewcommand{\bk}{\ensuremath {\boldsymbol {k}}}

\renewcommand{\bu}{\ensuremath {\boldsymbol {u}}}

\renewcommand{\bbf}{\ensuremath {{\boldsymbol {f}}}}
\newcommand{\bnabla}{\ensuremath {\boldsymbol {\nabla}}}

\newcommand{\bbeta}{\boldsymbol \beta}

\newcommand{\rdip}{\text{rD}}
\newcommand{\rstrat}{\text{rS}}

\begin{document}

\title{Mixing by microorganisms in stratified fluids}
\author{Gregory L. Wagner$^1$, William R. Young$^2$, and Eric Lauga$^{3}$\\
$^1$ Department of Mechanical and Aerospace Engineering, \\ University of California, San Diego, \\ 9500 Gilman Drive, La Jolla CA 92093-0411, USA;  \\
$^2$ Scripps Institution of Oceanography \\
9500 Gilman Drive,  La Jolla, CA 92093-0213, USA; \\
$^3$ Department of Applied Mathematics and Theoretical Physics, \\ 
University of Cambridge, 
Wilberforce Road, 
Cambridge, 
CB3 0WA, 
UK.}
\date\today

\maketitle

\begin{abstract}

We examine the vertical mixing induced by the swimming of microorganisms at low Reynolds and P\'eclet numbers in a stably stratified ocean, and show that the global contribution of oceanic microswimmers to vertical mixing is negligible. We propose two approaches to estimating the mixing efficiency, $\eta$, or the ratio of the rate of potential energy creation to the total rate-of-working on the ocean by microswimmers. The first  is based on scaling arguments and  estimates $\eta$ in terms of the ratio between the typical  organism size, $a$, and an intrinsic length scale for the stratified flow,  $\ell = \left ( \nu \kappa / N^2 \right )^{1/4}$, where $\nu$ is the  kinematic viscosity, $\kappa$ the  diffusivity, and $N$ the buoyancy frequency. In particular, for small organisms in the relevant oceanic limit, $a / \ell \ll 1$, we predict the scaling $\eta \sim (a / \ell)^3$. The second  estimate of $\eta$ is formed by solving the full coupled flow-stratification problem by modeling the swimmer as a regularized force dipole, and computing the efficiency numerically. Our computational results, which are examined for all ratios  $a/\ell$,   validate the scaling arguments in the limit $a / \ell \ll 1$ and  further predict $\eta \approx 1.2 \left ( a / \ell \right )^3$ for vertical swimming and $\eta \approx 0.15 \left ( a / \ell \right )^3$ for horizontal swimming.  These results,  relevant for any stratified fluid rich in biological activity,  imply that the mixing efficiency of swimming microorganisms in the ocean is at very most 8\% and is likely smaller by at least two orders of magnitude.  

\end{abstract}

\section{Introduction}

Vertical mixing, or the vertical transport of convected quantities like temperature and salt, is of fundamental importance to general circulation, climate, and life in the ocean \citep{Munk:1966p11604}.  It is therefore essential to identify the mechanisms which drive vertical mixing in order to predict, for example, the consequences of changed environmental and climactic conditions on ocean circulation and ecosystems \citep{Wunsch:2004p11597}.  Vertical mixing in the ocean can be assessed either by an effective or ``eddy'' diffusivity of convected quantities, or by the mixing efficiency of a forcing process such as tidal forcing, wind stress or, as we focus on in this paper, the swimming of organisms. The mixing efficiency, $\eta$, of a process in a stably stratified fluid like the ocean is defined as the ratio between the  rate-of-creation of gravitational potential energy and the total rate-of-working on the fluid, the rest being  turned into heat by viscous dissipation.  Non-zero values for $\eta$ are possible only in a stably stratified fluid: if the fluid is unstratified then in statistically steady situations all of the external work is dissipated by viscous stresses.  It is a fundamental insight of \cite{Munk:1966p11604} and  \cite{Munk:1998p11603} that the strong stable stratification of the ocean implies vertical mixing is limited by the power supply, and that top-down energy budgets strongly constrain eddy diffusivities. The mixing efficiency is a fundamental ingredient in these arguments. For example,  Osborn's (1980) inequality, $K_{\rho} < 0.2 \varepsilon/N^2$, for effective or eddy diffusivity $K_{\rho}$, buoyancy frequency $N$, and rate of viscous dissipation $\varepsilon$, corresponds to a value $\eta = 1/6$.

One vertical mixing mechanism receiving recent attention is the swimming of organisms. The geophysical significance of pelagic bioturbation can be argued using either energy production and transfer in the ocean biosphere or oxygen consumption; both arguments lead to an estimate of about 1 terawatt (TW) of total mechanical energy transfer to the deep ocean by swimming organisms \citep{Dewar:2006p11606}.  Bolstering this conjecture are predictions \citep{Huntley:2004p11615} and observations \citep{Kunze:2006p11622, Gregg:2009p11598} which find that  kinetic energy dissipation within aggregations of swimmers can reach $10^{-6}$ -- $10^{-5} \units{W/kg}$, much greater than the typical deep ocean rates of $10^{-9}$--$10^{-8} \units{W/kg}$.  But despite these significant levels of dissipation, scaling arguments \citep{Visser:2007p11607, Kunze:2011p11599} and a small number of observations \citep{Gregg:2009p11598} suggest that the characteristic lengths of biogenic eddies are too small for mixing efficiencies to be significant, and that most of the energy is therefore dissipated by viscous stress rather than stored in gravitational potential energy. In other words,  $\eta$ is too small for biogenic mixing to matter.   Nevertheless, it is possible that (a) energy transfer in aggregations may take place at scales larger than that of an individual swimmer, (b) non-turbulent transport mechanisms are important, and (c) mixing efficiency depends on the direction of swimming: in particular  one would expect intuitively that vertical swimming produces the largest $\eta$ \citep{Gregg:2009p11598,Dabiri:2010p11624}.

In this work, we consider the potential for mixing by a previously ignored source: microorganisms swimming at low Reynolds numbers, or Reynolds numbers which are much less than 1.  This class of swimmers includes bacteria and small planktonic organisms, and excludes larger zooplankton and copepods which swim with Reynolds numbers close to 1 or greater.   Despite their small size, it seems reasonable to consider the contribution of microorganisms to ocean mixing due to their great numbers and the fact that they constitute the bulk of biomass in the ocean \citep{Stocker:2012p11775}.  In the deep ocean (where their impact on vertical diffusivity might be most important) \cite{Whitman:1998p11623} estimate their average concentration at 50,000 individuals per$\units{cm^3}$; in the upper $200 \units{m}$ of the ocean this number is greater by an order of magnitude at $5 \times 10^{5} \units{cells/cm^3}$.  Their energy content is also large: using an approximate average Oxygen Utilization Rate (OUR) along with the physics of respiration, the total metabolic rate of bacteria has been estimated at $6 \units{TW}$ in the deep ocean \citep{Dewar:2006p11606}.  Furthermore, there is evidence that the ability to swim is widespread among bacteria: though the fraction of bacteria that swim may range from 5 to 70\% and is subject to large natural variability, the swimming fraction may be as high as 80\% in the presence of enhanced nutrient concentration \citep{Stocker:2012p11775}.  

Mixing by organisms in the low Reynolds number  has been studied extensively both theoretically \citep{Lin:2011p11654, Kunze:2011p11599} and experimentally \citep{Wu:2000p11649, Leptos:2009p11636}, with focus on the effective diffusivity induced in suspensions of microswimmers.  In this paper  we provide a complementary approach and  quantify mixing by microorganisms through their mixing efficiency in the low Reynolds number ($\Re$), low P\'eclet number limit ($\Pe$), which is the relevant one for the convection of temperature or salt by bacteria.  We first use scaling arguments to estimate the efficiency as a function of the typical ratio between the microorganism size and the intrinsic length scale in the stratified fluid. We then solve the full coupled flow-stratification problem by modeling the microorganism as a regularized force dipole, and evaluate the mixing efficiency numerically. These results validate our scaling approach and demonstrate that the mixing efficiency of a population of microorganisms can reach 8\% for microorganisms which are of similar size as the stratification length scale, but is on the order of 0.01\% for microorganisms and stratification levels relevant to the ocean, and thus negligible.  This conclusion confirms that \cite{Dewar:2006p11606} were correct in  excluding bacteria from their assessment of the total contribution of swimming organisms to the mechanical energy budget of the ocean.  Thus the major open question of biogenic mixing is the mixing efficiency of larger swimmers.

Our paper is organized as follows.  In \S\ref{setup} we derive the governing equations for fluid motion in the low Reynolds number, low P\'eclet number limit and introduce our model for a swimmer in this regime.  In \S\ref{energetics} we derive the mechanical energy equation and define the mixing efficiency $\eta$ as the ratio between the creation of gravitational potential energy and the rate-of-working on the fluid.  In \S\ref{scalingSec} we develop a scaling argument for mixing efficiency in low Reynolds and P\'eclet number flows which applies both to settling particles and swimming microorganisms which are much smaller than the stratification length scale.  In \S\ref{full} we present the results for mixing efficiency as estimated by our model, and discuss our findings in \S\ref{discussion}  In Appendix~\ref{appendixScalings} we present scaling arguments for the mixing efficiency of settling particles and microorganisms which are much larger than the stratification length scale.  In Appendix~\ref{appendixIntegrals} give the the integrals quantifying the rate of mechanical energy transfer to the fluid and to gravitational potential energy.  In Appendix~\ref{appendixEnsemble} we outline the averaging procedure used to estimate the mixing efficiency within a dilute ensemble of swimmers with uniformly distributed random orientations.  Finally in Appendix~\ref{appendixAsymptotics} we discuss the asymptotic evaluation of the mechanical energy integrals in the two limiting cases where the microorganism is either much smaller or much larger than the intrinsic stratification length scale.

\section{Governing equations for stratified locomotion by microorganisms}
\label{setup}

In order to quantify the mixing of stratified fluids by swimming microorganisms we first derive a simple system of equations which models  the action of the microorganism on the fluid as a force density, $\bbf$, spatially distributed in the fluid (but with no net force).  Of critical importance to this derivation is the Reynolds number (denoted $\Re$), or the ratio between inertial and viscous forces, and the P\'eclet number (denoted $\Pe$), or the ratio between advection and diffusion in the transport of either temperature or salt.  For example, the marine bacterium {\it Pseudoalteromonas haloplanktis} has a size on the order of $a \approx 1 \units{\mu m}$ and swims at speeds around $U = 80 \units{\mu / s}$, which implies that $\Re = U a / \nu \approx 8 \times 10^{-5}$ \citep{Stocker:2012p11775}.  If {\it P. haloplanktis} is reasonably representative of deep sea marine bacteria, then because $\Pe_{\text{temp}} \approx  7 \,\Re$ for temperature stratification and $\Pe_{\text{salt}} \approx 700\, \Re$ for salt stratification, the locomotion of most marine bacteria is associated with both  low $\Re$ and low $\Pe$.  

\subsection{Dynamics of a forced stratified fluid}

To begin we write the the fluid density, $\rho$, as the sum of three contributions: a reference density, $\rho_0$, a background density gradient, and perturbations from this background density gradient expressed in terms of the buoyancy, $b$,  representing the acceleration imparted to fluid elements due to this deviation,
\begin{equation}
\rho = \rho_0 \left [ 1 - g^{-1} \left ( N^2 z + b \right ) \right ],
\end{equation}
where $g$ is gravitational acceleration, $N = \sqrt{-\left ( g / \rho_0 \right ) \partial \rho / \partial z}$ is the  buoyancy frequency, and the $z$-coordinate is aligned with gravity.  
If we assume that the background gradient and perturbation introduce only small deviations from the reference density $\rho_0$, we may make the Boussinesq approximation and write the Navier-Stokes equations in the form
\begin{eqnarray}
\nabla \cdot \bu &= &0, \\
\rho_0 \left ( \frac{\partial \bu}{\partial t} + \bu \cdot \nabla \bu \right ) &= &- \nabla p + \rho_0 b \, \bzh + \mu \nabla^2 \bu + \bbf\, , \label{mom1}
\end{eqnarray}
where $\bu = \{u, v, w\}$ is the velocity field in the fluid, $p$
  the disturbance pressure, $\bzh$ is a unit vector in the vertical $z$-direction,  $\mu$ the dynamic viscosity of the fluid, and $\bbf(\bx)$ is a body force density (with dimensions of force per unit volume)  used to model the flow disturbance induced in the fluid by the microswimmer.

We assume that the fluid density is determined by a single stratifying agent like salt or temperature through a linear equation of state.  The distribution of density is then governed by an advection-diffusion equation which can be expressed in terms of the buoyancy $b$ as
\begin{equation}
\frac{\partial b}{\partial t} + \bu \cdot \nabla b + w N^2 = \kappa \nabla^2 b\, ,
\label{buoy1}
\end{equation}
where $\kappa$ is the molecular diffusivity of the stratifying agent and the  term $w N^2$ arises from advection of the background density gradient by vertical fluid motion. 

\subsection{Modeling swimmers as a regularized force dipole}

Various modeling approaches can approximate the flow induced by a  swimming microorganism \citep{brennen77,Powers:2009p11653}.  
The most detailed models require realistic geometry and a deformable boundary.   A simpler possibility is the ``squirmer''  proposed by Lighthill, which models a swimming microorganism as a spherical body with a tangential velocity distribution imposed on its surface \citep{Lighthill1952,Blake1971b}. An even more idealized  model  is the representation of the microswimmer by a dipolar force singularity.  For unstratified Stokes flow, whose governing equations are linear, the properties of solutions forced by singularities -- the Green's functions of the Stokes equation and its derivatives --  are well-known \citep{Chwang:1975p11746,kim2005}.  The velocity distribution of a force dipole singularity, representing the simultaneous and opposite action of the propelling flagella and the drag of the microorganism, has been shown to correspond well to the flow field generated by a single bacterium \citep{Drescher:2011p11619}. Models where swimmers are approximated as force-dipoles have been successful in reproducing some of the behaviors and characteristics peculiar to self-propelled microorganisms \citep{Powers:2009p11653}.  

However, modeling a swimmer as a force-dipole leads to a mathematical singularity and infinite viscous dissipation. This unphysical result indicates that information about the size of the microorganism is  essential to any attempt at estimating the mixing efficiency. In this work we take inspiration from the singularity model but smooth or ``regularize'' the singularity \citep{cortez01}. Specifically, we replace  the $\delta$-function in the usual Green's function by a Gaussian.  This distributes the total forcing over a finite region of fluid, which we  identify with the characteristic size, $a$, of the microorganism.  Such regularized singularities have long been used to obtain   efficient solution of the boundary integral formulation of the Stokes equations \citep{Cortez:2005p11613}.  

The regularized point force, or regularized ``Stratlet'' in the context of stratified fluids  \citep{List:1971p11611, Ardekani:2010p11593}, corresponds to a force density given by 
\beq
\bbf_{\text{reg. Stratlet}}  =  \frac{e^{-r^2 / 2 a^2}}{\left ( \sqrt{2 \pi a} \right )^3} 
\,\label{F1} \bF,
\end{equation}
where the constant vector $\bF$ indicates the direction and magnitude of the total force acting on the fluid.  The regularization is such that the forcing $\bbf$ limits to a delta function as $a \to 0$ and the corresponding solution limits  to the Green's function.  Here we use a convenient Gaussian form for the ``cut-off'' function, though any function which limits to a delta function will work \citep{cortez01}.   

The regularized dipole is then derived from the regularized Stratlet solution,
\begin{equation}
\bbf_{\text{reg. dipole}} = - \bbeta \cdot \nabla \bbf_{\text{reg. Stratlet}} = - \bF \bbeta \cdot \bnabla \frac{e^{-r^2 / 2 a^2}}{\left ( \sqrt{2 \pi a} \right )^3} ,
\label{F2}
\end{equation}
where $\bbeta$, with  dimensions of length, is the displacement between the two constituent point forces in the dipole.  For an organism, these two point forces correspond  to the equal and opposite forces exerted by the organism body on the fluid and by the action of the flagella on the fluid, so that the total force is zero.  As such $\bbeta$ is always either parallel or anti-parallel to $\bF$ and its magnitude $| \bbeta |$ roughly corresponds to the size of the organism \citep{Powers:2009p11653}.  We define the total magnitude of the dipole to be $D = | \bbeta | | \bF |$, which has dimensions of force $\times$ length.  

\subsection{Non-dimensionalization}

We proceed with the derivation of governing equations by scaling the equations and analyzing the relative magnitude of each terms. A primary external parameter is the magnitude of the  force $F =|\bF|$, appearing in the dipole.  For bacterium, the typical order of magnitude of the propulsive force of a flagellum is $F \sim 10^{-12} \units{N}$ \citep{Drescher:2011p11619}. We  introduce a characteristic length scale, $L$, that may be thought of as the size of the microswimmer, denoted by  $a$ in \eqref{F1} and \eqref{F2}, or alternately as a length scale for induced fluid motions (which may different than $a$ for an ensemble of microorganisms). With the low Reynolds number limit in mind, we introduce the velocity
\begin{equation}
U \equiv  \frac{F }{ \mu L}, \label{Udef}
\end{equation}
which is the typical swimming velocity of the microorganism when the distributed propulsive force $F$ is balanced by viscous stresses in a domain of size $L$. Using $U$, $F$ and $L$, we then scale the equations of motion as
\begin{gather}
\bx = L \bx'\, , \qquad \qquad  t = \frac{L}{U} t'\, , \qquad \qquad \bbf = \frac{F}{L^3} \bbf'\, , \\
\bu = U \bu'\, , \qquad \quad p = \frac{F}{L^2} p'\, , \qquad \quad b = \frac{U N^2 L^2}{\kappa} b'\, ,
\end{gather}
where a prime denotes a non-dimensional variable.  The scaling for buoyancy arises from assuming a balance between the advection of the background gradient, $N^2$, and diffusion of buoyancy, and may be viewed alternatively as an assumption that $b$ is an $O(\Pe)$ correction to the background buoyancy field, $N^2 z$.    These scalings yield the non-dimensional system
\begin{eqnarray}
\nabla \cdot \bu' &= &0, \label{nondimmom1}\\
\Re \left ( \frac{\partial \bu'}{\partial t'} + \bu' \cdot \nabla \bu' \right ) &=& - \nabla p' + \left ( \frac{L}{\ell} \right )^4 b' \, \bzh + \nabla^2 \bu' + \bbf', \label{nondimmom2} \\
\Pe \left ( \frac{\partial b'}{\partial t'} + \bu' \cdot \nabla b' \right ) + w' &=& \nabla^2 b',
\label{nondimmom3}\end{eqnarray}
where the Reynolds and  P\'eclet numbers are
\beq
\Re \equiv \frac{F}{ \rho_0 \nu^2} , \quad  \Pe \equiv \frac{F}{ \rho_0 \nu \kappa},
\eeq
with  $\nu= \mu/\rho_0$  the kinematic viscosity.  
The length $\ell$ appearing in \eqref{nondimmom2} is the intrinsic stratification length scale given by 
\begin{equation}
\ell \equiv \left ( \frac{\nu \kappa}{N^2} \right )^{1/4} .
\label{elldef}
\end{equation}
In addition to the two expected dimensionless parameters, $\Re$ and  $\Pe$, we see that  the ratio of length scales,  $\left ( L / \ell \right )^4$, multiplies the buoyancy term in the fluid momentum equation   \eqref{nondimmom2} and its magnitude thus determines the  importance of buoyancy forces.    

What is the typical value for $\ell$? To derive an estimate we must consider not only the overall stratification of the ocean, but the microstructure and small-scale variation in stratification which result from turbulent motions and disordered displacements of fluid on the scale of the microorganism.  If we assume that the small scale variability in the temperature and salinity gradient is at most about  200 times the mean value \citep{Gregg:1977p11639}, and that $N$ in the ocean measured on length scales of tens of meters varies roughly from 0.2 to 4 {cycles/hour}  \citep{talley2011}, then we find that $\ell \approx 100 \units{\mu m}$ to $10 \units{mm}$ for salt stratification and $\ell \approx 500 \units{\mu m}$ to $40 \units{mm}$ for thermal stratification.  Most swimming microorganisms in the ocean  are bacteria and plankton with a typical size ranging from $ 1 \units{\mu m}$ 
to $ 100 \units{\mu m}$, which implies that in regions of strong local stratification the grouping $\left ( L / \ell \right )^4$ can be at most order one, and otherwise is typically very small.

\subsection{Leading-order linear system of equations at low $\Re$ and $\Pe$}

If we retain only the leading-order terms in $\Re$ and $\Pe$ in \eqref{nondimmom1} -- \eqref{nondimmom3}, as is appropriate for microorganisms in either temperature or salt stratification, and restore the dimensionality of the equations, we obtain a linear system of equations describing the  motion of the stratified fluid driven by a force density, $\bbf$, at low Reynolds and P\'eclet number,
\begin{eqnarray}
\nabla \cdot \bu &=& 0, \label{lowRePe_mass}   \\
\nabla p - \mu \nabla^2 \bu &=& \rho_0 b \, \bzh + \bbf, \label{lowRePe_mom} \\
w N^2 &=& \kappa \nabla^2 b,
\label{lowRePe_buoyancy}
\end{eqnarray}
where the force density for our model  swimmer is given by \eqref{F2}. The linearity of these equations will allow us to calculate their solutions analytically using Fourier transforms.

The fundamental solution to this system of equations, or the solution corresponding to a point force $\bbf = \bF \delta (\bx)$, was first analyzed by \cite{List:1971p11611} and later termed the ``Stratlet'' by \cite{Ardekani:2010p11593}, with reference to the ``Stokeslet'' solution for a point force in unstratified low Reynolds number flow.  As would be expected, both \cite{List:1971p11611} and \cite{Ardekani:2010p11593} found that vertical fluid motion is suppressed by stratification.  For example, \cite{List:1971p11611} showed that for the horizontally-oriented Stratlet the vertical velocity decays exponentially for $z \gg x, y$, where $z$ is the direction of straitification, and as $s^{-7/3}$ for large $s$, where $s = \sqrt{x^2 + y^2}$ is the distance from the singularity in the horizontal plane $z=0$.  This contrasts with the Stokeslet, for which all velocities decay with $1 / r$, where $r$ is the distance to the singularity.

\section{Mixing efficiency}
\label{energetics}

The mechanical energy equation is derived by taking the dot product of the momentum equation \eqref{mom1} with the  velocity field,  $\bu$, and integrating over all of space. Using the divergence theorem, and assuming that the disturbance decays sufficiently quickly as  $|\bx| \gg 1$, one finds the kinetic energy equation,
\begin{equation}
\frac{\d}{\d t} \int \tfrac{1}{2} \rho_0 |\bu|^2 \, \d V + \underbrace{ \mu \int  |\nabla \bu|^2  \, \d V}_{\equiv P_{\text{visc}}}  \underbrace{ - \rho_0 \int  w b \, \d V}_{\equiv P_g} = \underbrace{\int \bu \!\cdot\! \bbf\, \d V }_{\equiv P_{\tot}}, 
\label{powerIntegralDefinitions}
\end{equation}
where we have defined three terms: $P_{\tot}$ is the rate at which the force density,  $\bbf$,  is working on the fluid,  $P_{\text{visc}}$ is the rate at which this work is turned into heat by viscous dissipation, and $P_g$ is the rate of creation of gravitational potential energy.  We can obtain another expression for $P_g$ by multiplying the buoyancy equation \eqref{buoy1} by $b$ and integrating over the volume, leading to
\begin{equation}\label{newb}
\frac{\d}{\d t} \int \tfrac{1}{2} b^2 \, \d V + N^2 \int w b \, \d V = - \kappa \int |\nabla b|^2\, \d V\, .
\end{equation}
We are concerned with steady flows  hence both ${\d}/{\d t}$ terms in \eqref{powerIntegralDefinitions} and \eqref{newb} disappear. Defining the mixing efficiency, $\eta$, as the ratio between ${P_g}$ and $P_{\tot}$, we obtain 
\begin{equation}
\eta = \frac{P_g}{P_{\tot}} =   \frac{ \rho_0 \kappa \int |\nabla b|^2 \d V}{N^2 \int \bu \!\cdot\! \bbf \d V}\cdot
\label{eta2}
\end{equation}
We always have  $\eta\leq 1$, and the larger the efficiency, the larger the proportion of the forcing used for mixing.  Note that following \cite{Osborn:1980p11625}, the mixing efficiency $\eta$  can be related to the ``flux coefficient'' $\Gamma = P_g / P_{\text{visc}}$ (the ratio between the rate of creation of gravitational potential energy and the rate of viscous dissipation) by $\eta = \Gamma / (\Gamma + 1)$.

\section{Scaling argument for small particles and swimmers in weak stratification \label{scalingSec}}

Let us consider a small settling particle, or a self-propelled microswimmer, with a characteristic size $a$. What are the expected scalings for both energetic contributions $P_{\tot} $ and $P_g$? The relevant limit to consider for microorganisms is $a\ll \ell$, which we consider below. The complementary  limit $a/\ell \gg1 $ is addressed in Appendix \ref{appendixScalings}.

The scaling for the total power input is  straightforward. Indeed $P_{\tot} $  should scale with the product of some characteristic fluid stress, $\sigma$, exerted by the fluid on the particle and some characteristic velocity, $U$, such that $P_{\tot} \sim \sigma  U a^2$.  In both cases of  settling particles and swimming at low Reynolds number, the flow equations are linear and thus we expect the typical stress to scale linearly with velocity with a viscous relationship $\sigma \sim \mu U/a$ and thus obtain the classical scaling for the total power input 
\begin{equation}
\label{Ptot}
P_{\tot} \sim \mu a U^2.
\end{equation}

Developing a scaling argument for the rate of increase of the gravitational potential energy requires a detailed look at the fluid dynamics.  At low Reynolds number the velocity disturbance far from a settling particle in an unstratified fluid looks like the disturbance due to a point force and decays as $1/r$, where $r$ is the distance from the particle. In contrast,  the velocity disturbance far from a swimmer at low Reynolds number looks like the disturbance created by a force dipole and decays as $1/r^2$ \citep{Chwang:1975p11746,Powers:2009p11653}.  This asymptotic difference is related to the fact that a settling particle exerts a net force on the fluid, while a microswimmer does not. Below we show that this results in quite different mixing efficiences: a settling particle is much more efficient than a microswimmer. However, in both cases, the algebraic decay  takes place only  within some region $a  \lesssim r \lesssim \ell$ where the buoyancy term in the momentum equation (\ref{lowRePe_mom}) is negligible, and beyond a length scale $O(\ell)$   the vertical velocity is suppressed by stratification.

For a settling particle the velocity then scales as $|\bu| \sim U (a / r)$, which we can then insert into the buoyancy conservation equation to find a scaling for the buoyancy,
\begin{equation}
U \left ( \frac{a}{r} \right ) N^2 \sim \kappa \nabla^2 b ,
\end{equation}
leading to
\begin{equation}
 b \sim \frac{U a r N^2}{\kappa}\cdot
\end{equation}
The rate of creation of gravitational potential energy is then given by an integral over a volume of size $O(\ell^3)$
\begin{equation}
P_g \sim - \rho_0 \int\!\!\!\int\!\!\!\int_{r<\ell} w b \d V 
\end{equation}
and thus we obtain the scaling relation
\begin{equation}
P_g\sim \rho_0 \int_0^{\ell} U \left ( \frac{a}{r} \right ) \frac{U a r N^2}{\kappa} r^2 \d r \sim \frac{\rho_0 U^2 a^2 N^2 \ell^3}{\kappa}\cdot\label{Pg_force}
\end{equation}
Since  the mixing efficiency is defined as $\eta = P_g / P_{\tot}$, we can use \eqref{Ptot} and \eqref{Pg_force} and recall the definition  $\ell = \left ( \nu \kappa / N^2 \right )^{1/4}$, to find that for $a / \ell \ll 1$, 
\begin{equation}
\eta  \sim  \displaystyle \frac{a}{\ell},
\end{equation}
for a settling particle which exerts a net force on the fluid. 

In the case of the microswimmer,  the typical velocity decays faster as $\bu \sim U (a / r)^2$, which implies $b \sim U a^2 N^2 / \kappa$ and $P_g \sim \rho_0 U^2 a^4 N^2 \ell / \kappa$ by the same argument given for the settling particle, leading to an efficiency  
scaling as 
\begin{eqnarray}
\eta  \sim  \displaystyle \left ( \frac{a}{\ell} \right )^3 ,
\end{eqnarray}
for a swimmer exerting a dipolar perturbation on the fluid. While this scaling analysis does not determine  multiplicative constants, it indicates however that mixing efficiencies by microorganisms  should be expected to be small. 
The smallest values of  $\ell$ in the ocean correspond to a strong salt stratification and are around $100 \units{\mu m}$. For larger microorganisms which have a size on the order of $10 \units{\mu m}$, we obtain an estimate for the efficiency of $\eta\sim 0.1\%$.

\section{Solving the full regularized model}
\label{full}
Moving beyond the scaling analysis, in this section we solve the governing equations analytically for the regularized microswimmer model.  As we detail below, the full solution recovers the physical scalings from the previous section and also determines the multiplicative constants. 
 
\subsection{Solution in Fourier space}
 
We employ the three-dimensional Fourier transform--inverse transform pair defined for an arbitrary function $g(\bx)$ as 
\begin{eqnarray}
\cF \left [ g(\bx) \right ] = \tilde g (\bk) &=& \int_{\R^3} g(\bx) e^{-i \bk \cdot \bx} \d V, \\
g(\bx) &=& \frac{1}{8 \pi^3} \int_{\ksp} \tilde g (\bk) e^{i \bk \cdot \bx} \d \bk,
\end{eqnarray}
where the $\tilde g (\bk) $ is the Fourier transform of $g(\bx)$ and $\bk = \{k_1, k_2, k_3\}$ is the wave number vector.  Applying this Fourier transform to  (\ref{lowRePe_mass}) -- (\ref{lowRePe_buoyancy}) and using the identity $\partial / \partial x_i \to i k_i$ yields the algebraic system
\begin{eqnarray}
\bk \cdot \tilde \bu &=& 0, \\
i \bk \tilde p + \mu k^2 \tilde \bu &=& \rho_0 \tilde b \, \bzh + \tilde \bbf, \label{ftMom} \\
\tilde w N^2 &=& - \kappa k^2 \tilde b. \label{ftBuoy}
\end{eqnarray}
This system is easily solved by taking the dot product between $\bk$ and \eqref{ftMom}, solving for the pressure, and then combining \eqref{ftBuoy} with the $\bzh$-component of \eqref{ftMom} to find the buoyancy.  The solution can be written concisely in spherical coordinates, where the physical space coordinates $\bx = \{r, \theta, \phi \}$ correspond to the Fourier space coordinates $\bk = \{k, \psi, \omega \}$.  In this case, the unit vectors $\hat{\bk} = \bk / k$ and $\hat{\bpsi}$ can be written in terms of the Cartesian unit vectors $\bxh$, $\byh$, and $\bzh$ as
\begin{eqnarray}
\hat{\bpsi} &=& \bxh \cos \psi \cos \omega + \byh \cos \psi \sin \omega - \bzh \sin \psi, \\
\hat{\bk} &=& \bxh \sin \psi \cos \omega + \byh \sin \psi \sin \omega + \bzh \cos \psi.
\end{eqnarray}
Inspection of these relations reveals that $\hat{\bk} \cos \psi = \hat{\bpsi} \sin \psi + \bzh$.  Armed with this identity, we find we can write the buoyancy simply as
\beq
\tilde b = \frac{\sin \psi}{\left ( k \ell \right )^4 + \sin^2 \psi} \; \hat{\bpsi} \cdot \left ( \tilde \bbf / \rho_0 \right ).
\eeq
The pressure then becomes
\beq
\tilde p = - \frac{i}{k} \left ( \hat{\bk} + \hat{\bpsi} \frac{\cos \psi \sin \psi}{\left ( k \ell \right )^4 + \sin^2 \psi} \right ) \cdot \tilde \bbf,
\eeq
and the fluid velocity is
\beq
\begin{split}
\tilde \bu &= \frac{1}{\mu k^2} \left ( \bI - \hat{\bk} \hat{\bk} \right ) \cdot \tilde \bbf - \hat{\bpsi} \, \tilde b \, \frac{\sin \psi}{\nu k^2} , \\
&= \frac{1}{\mu k^2} \left ( \bI - \hat{\bk} \hat{\bk}  - \hat{\bpsi} \hat{\bpsi} \frac{\sin^2 \psi}{\left ( k \ell \right )^4 + \sin^2 \psi} \right ) \cdot \tilde \bbf .
\end{split}
\eeq
Written in this form, we see clearly the part of the solution $\tilde \bu$ which corresponds to the solution for unstratified low Reynolds number flow (to which this solution limits when $b =0$), and the part which corresponds to buoyancy forces induced by the presence of a density stratification.  This solution is identical to the solution for a point force in stratified low Reynolds number flow, or the ``Stratlet'' \citep{List:1971p11611, Ardekani:2010p11593} with our general forcing term $\tilde \bbf$ replacing the point force. In Fourier space, the regularized dipole is then easily derived from the solution for the regularized Stokeslet as
\begin{equation}
\{ \tilde \bu, \tilde p, \tilde b, \tilde \bbf \}_{\text{\rdip}} = - i \left ( \bbeta \cdot \bk \right ) \{ \tilde \bu, \tilde p, \tilde b, \tilde \bbf\}_{\text{\rstrat}}\, .
\label{dipole_F}
\end{equation}
The formulas above provide the full analytical solution to the low--$\Re$, low--$\Pe $ problem in Fourier space.

 \subsection{Solution in physical space}

\begin{figure}[t]
\centering
\includegraphics[width = .95\textwidth]{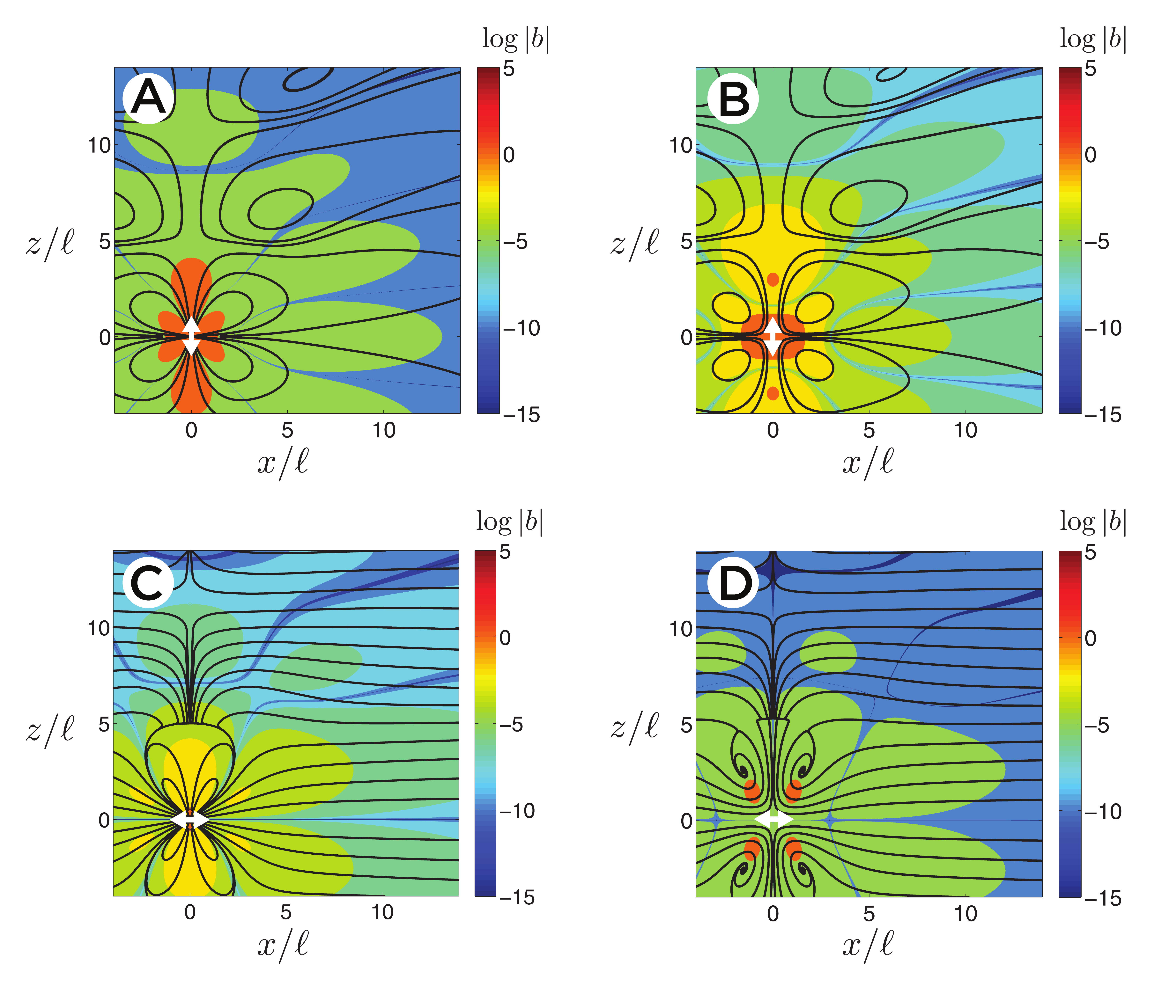}
\caption{Buoyancy isocontours and streamlines for (A) the vertically-oriented Stratlet dipole, (B) the vertically-oriented regularized Stratlet dipole with $a / \ell = 1/2$, (C) the horizontally-oriented Stratlet dipole, and (D) the horizontally-oriented regularized Stratlet dipole with $a / \ell = 1/2$.}
\label{solutions}
\end{figure}

In order to visualize the solution we can invert the Fourier space solutions back into physical space numerically by adapting MATLAB's $\texttt{ifft}$ function. A contour plot of isocontours of buoyancy with streamlines superimposed is shown in Fig.~\ref{solutions} for $a=0$ (Stratlet dipole, Fig.~\ref{solutions}A and C) and $a = 1/2$ (regularized Stratlet dipole, Fig.~\ref{solutions}B and D) for a vertical and horizontally-oriented microswimmer ($\bbeta = | \bbeta | \bzh$ and $\bxh$, respectively).

In all four cases, the flow field near the microswimmer displays the usual dipolar structure while farther away (roughly at a distance $r \approx \ell$) the vertical flow is suppressed by the stratification.  For the vertically-oriented swimmers, the flow field is organized into toroidal recirculation regions which extend through the entire domain.  The horizontally-oriented microswimmers, on the other hand, induce only a small recirculation region close to the origin, and the flow field is essentially horizontal when $r / \ell \approx 10$.  The effect of the regularization is to weaken the flow field and to move recirculating regions  away from the origin.

 \subsection{Mixing efficiency}

The values of  $P_g$ and $P_{\tot}$, and thus the mixing efficiency, can be evaluated directly in Fourier space  using Parseval's theorem to relate  integral expressions over all physical space to integrals over Fourier space. Specifically we have
\begin{equation}
P_g = - \rho_0 \int_{\R^3} w(\bx) b(\bx) \d V = - \frac{\rho_0}{8 \pi^3} \int_{\ksp} \!\!\!\!\!\! \tilde w(\bk) \tilde b(-\bk) \d k,
\end{equation}
and
\begin{equation}
P_{\tot} = \int_{\R^3} \bu(\bx) \cdot \bbf(\bx) \d V = \frac{1}{8 \pi^3} \int_{\ksp} \!\!\!\!\!\!  \tilde \bu(-\bk) \cdot \tilde \bbf(\bk) \d k.
\end{equation}

We consider first the case of a microswimmer oriented vertically, so that the solution is axisymmetric. If we then insert the regularized dipole solution, express the integral in spherical coordinates $\{k, \psi, \theta\}$ with $k_1 = k \cos \theta \sin \psi$, $k_2 = k \sin \theta \sin \psi$, $k_3 = k \cos \psi$, and $k = | \bk |$, non-dimensionalize the integral using $k' = k \ell$ and drop the primes for simplicity, and integrate over $\theta$, we obtain the expressions
\begin{equation}
P_g = \frac{D^2}{4 \mu \ell \pi^2} \int_0^{\infty} V(k) e^{- ( a k / \ell)^2} \d k, 
\label{vert_Pg}
\end{equation}
and
\begin{equation}
P_{\tot} = \frac{D^2}{4 \mu \ell \pi^2} \int_0^{\infty} W(k) e^{- (a k / \ell)^2} \d k,
\label{vert_Ptot}
\end{equation}
where   $D = | \bbeta | | \bF |$ is the total magnitude of the regularized dipole.  The functions $V(k)$ and $W(k)$ are defined by
\begin{equation}
\begin{split}
V(k) &= k^6 \int_0^{\pi}  \frac{\cos^2 \psi \sin^5 \psi}{\left ( k^4 + \sin^2 \psi \right )^2} \d \psi, \\
&= \frac{1}{3} k^6 \left ( 2 + 15 k^4 \right ) + \frac{k^{10}}{\sqrt{1+k^4}} \left ( 4 + 5 k^4 \right ) \log \left [ \frac{1}{k^2} \left ( \sqrt{1+k^4} - 1\right ) \right ],
\end{split}
\label{vk}
\eeq
and
\beq
\begin{split}
W(k) &= k^6 \int_0^{\pi} \frac{\cos^2 \psi \sin^3 \psi}{k^4 + \sin^2 \psi} \d \psi, \\
&= \frac{2}{3} k^6 \left ( 1 + 3 k^4 \right ) + 2 k^{10} \sqrt{1+k^4} \log \left [ \frac{1}{k^2} \left ( \sqrt{1+k^4} - 1\right ) \right ].
\end{split}
\label{wk}
\eeq
We find similar expressions for the horizontally-oriented dipole and slightly more complicated expressions for an arbitrarily oriented dipole;  both are given in Appendix~\ref{appendixIntegrals}.  

The integrals can be analyzed numerically and we find they are also amenable to asymptotic analysis in the limits where $a / \ell$ is either large or small.  The details of these asymptotic analyses are given in Appendix \ref{appendixAsymptotics}.   In the limit $a / \ell \ll  1$, which is most relevant to microorganisms in the ocean, we find for vertically and horizontally microswimmers
\begin{equation}
\eta_{\text{vert}}\, (a / \ell \ll 1) = 1.21 \left ( \frac{a}{\ell} \right )^3 - 2.29 \left ( \frac{a}{\ell} \right )^4 
+ O  \left ( \frac{a}{\ell} \right )^5,
\end{equation}
and
\begin{equation}
\eta_{\text{horz}}\,(a / \ell \ll 1) = 0.151 \left ( \frac{a}{\ell} \right )^3 - 0.286 \left ( \frac{a}{\ell} \right )^4 + O  \left ( \frac{a}{\ell} \right )^5 .
\end{equation}
These asymptotic results confirm the prediction of the scaling analysis, $\eta\sim (a/\ell)^3$,  and further show that in this limit,  the mixing efficiency of a horizontal microswimmer is about one order of magnitude below that of a vertical microswimmer.

In order to calculate the mixing efficiency for all values $a / \ell$, we  numerically compute the integrals in  \eqref{vert_Pg}, \eqref{vert_Ptot}, \eqref{horz_Pg}, and \eqref{horz_Ptot}.  The results are plotted in Fig.~\ref{mixingEfficiency} where we show the mixing efficiency as a function of the ratio $a / \ell$ for microorganisms swimming vertically (blue solid line) and horizontally (red dashed line). We also consider the more relevant case of  an ensemble of microorganisms whose orientations are uniformly distributed (black dotted line). The mathematical averaging for an ensemble is detailed in Appendix~\ref{appendixEnsemble}.  Briefly, the calculation can be reduced to a weighted average of contributions for the respective total work and gravitational potential energy terms from the horizontally-oriented force dipole, vertically-oriented force dipole, and contributions from additional singularities.  

\begin{figure}[t]
\centering
\includegraphics[width = .95\textwidth]{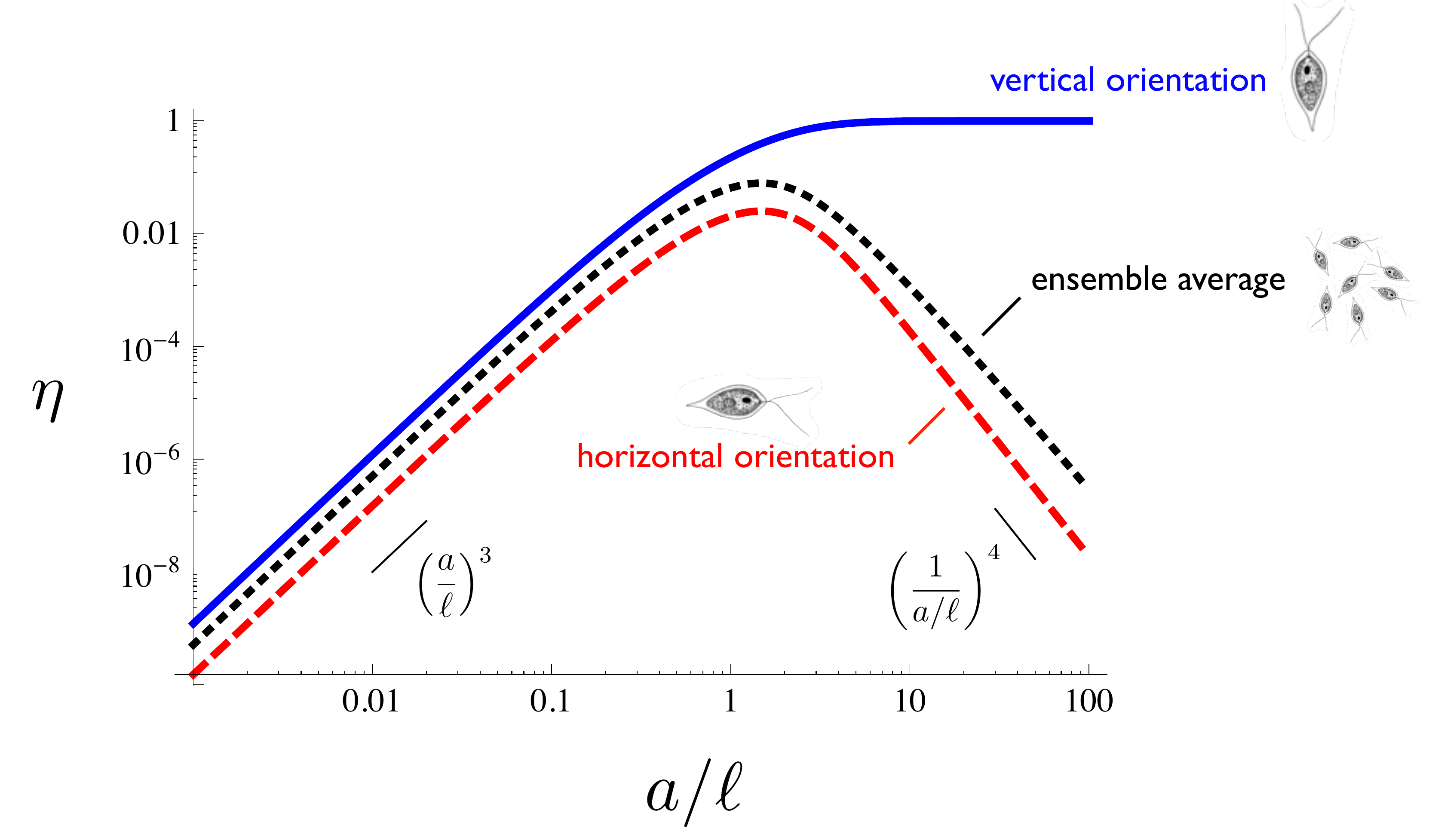}
\caption{Mixing efficiency for the regularized Stratlet dipole, 
$\eta$,  as a function of the length ratio $a / \ell$ for vertical microswimmers (blue, solid), horizontal microswimmers (red, dashed), and an ensemble average over a uniform distribution of swimmer orientations (black, dotted).  All three mixing efficiencies scale as $\left ( a / \ell \right )^3$ when $a / \ell \ll 1$.  The mixing efficiency of the vertically-oriented regularized Stratlet dipole asymptotes to 1 when $a / \ell \gg 1$ (see Appendix \ref{appendixScalings}) whereas for a  horizontally-oriented regularized Stratlet dipole it decreases as $\left ( a / \ell \right )^{-4}$ when $a / \ell \gg 1$.  
The ensemble average should approach $\left ( a / \ell \right )^{-4}$ as well for $a / \ell \gg 1$ though the approach is slow within the range shown here.}
\label{mixingEfficiency}
\end{figure}

We first find that for $a / \ell \ll  1$, the regularized singularity model confirms the scaling $\eta \sim (a / \ell)^3$ and agrees with the asymptotic result.  When $a / \ell \gg 1$, the mixing efficiency predicted by the distributed force model for the vertically-oriented  swimmers tends to 1.  Physically,  the stratification is being lifted directly by the distributed force.  Mathematically, it is a consequence of a dominant balance in the $z$-momentum equation between the distributed forcing and buoyancy.  In contrast, for the horizontally-oriented microswimmer, mixing efficiency decays as $\eta \sim \left(a / \ell \right)^{-4}$ because the fluid motion incurred by the regularized force is increasingly two-dimensional as $a / \ell$ increases.  In Appendix~\ref{appendixScalings}  we present a scaling argument to  explain this behavior of the  mixing efficiency.  In the ensemble average, the mixing efficiency approaches $\left (a / \ell \right )^{-4}$ as in the horizontal case.  These results can also be predicted by analyzing the integrals asymptotically, as shown in  Appendix~\ref{appendixAsymptotics}.  

Finally, we calculate the mixing efficiency expected from an ensemble of randomly oriented, non-interacting microorganisms.  The average total rate-of-working, $\Pavg_{\tot}$, and rate of creation of gravitational potential energy, $\Pavg_g$, are calculated as
\beq
\Pavg_{g, \tot} = \frac{1}{4 \pi} \int_0^{2 \pi} \int_0^{\pi} P_{g,\tot}(\alpha) \sin \alpha \d \alpha \d \theta = \frac{1}{2} \int_0^{\pi} P_{g,\tot}(\alpha) \sin \alpha \d \alpha,
\eeq
where $P_{g, \tot}(\alpha)$ is the total rate-of-working or rate of creation of gravitational potential energy for a swimmer oriented at an angle $\alpha$ with respect to vertical, and $\theta$ is the azimuthal angle.  From these we calculate the mixing efficiency of an ensemble as $\overline{\eta} = \Pavg_{g} / \Pavg_{\tot}$, plotted as the black dotted line in Figure~\ref{mixingEfficiency}.  We find that the mixing efficiency of the ensemble reaches a maximum of $\overline{\eta} = 8\%$ near $a / \ell \approx 1$, providing an upper bound on the mixing efficiency of a dilute suspension of microorganisms where correlations between the motion of individuals can be neglected.

\section{Discussion}
\label{discussion}

We have used  scaling arguments and the solution of the  Stokes equation in a stratified fluid to find the relationship between organism size,  fluid properties, stratification, and  mixing efficiency for small microorganisms associated with Reynolds numbers much less than 1.  The length $\ell$ in \eqref{elldef} combines all the relevant environmental properties of the fluid, and the ratio of the organism size, $a$, to the ``stratification length", $\ell$, is the crucial control parameter that determines mixing efficiency in the visco-diffusive regime.  

The strength of the scaling argument  is that it is independent of the particular form of the model, relying only on the (universal) dipolar nature of the flow field around the swimming microorganism when $a / \ell \ll 1$.  The scaling argument also yields physical insight: when $a / \ell \ll 1$, the total rate-of-working on the fluid is largely dissipated by viscous stress in a region comparable to the size of the microorganism, whereas gravitational potential energy is created within a larger region of size $\ell$.  An important consequence of this physical picture for $a / \ell \ll 1$ is that the mixing efficiency depends on the spatial structure of the induced velocity disturbance. Specifically,  for force-free swimmers where velocities decay like $1/r^2$, it scales as $\eta \sim ( a / \ell )^3$, whereas for sinking particles or rising bubbles exerting a net force on the fluid, and where velocities decay like $1/r$, it scales as $\eta \sim a / \ell$.  When $a / \ell \gg 1$, the scaling analysis detailed in  Appendix \ref{appendixScalings} implies that the velocity field resulting from a horizontal disturbance is largely two-dimensional, as intuitively expected.

The physical picture implied by this scaling analysis differs substantially from that implied by the scaling analysis for eddy diffusivities developed in \cite{Kunze:2011p11599}, which does not depend on the monopolar or dipolar nature of the velocity field induced by settling particles or force-free swimmers, respectively.  Additionally, \cite{Kunze:2011p11599} found that the effective diffusivity has no dependence on the scalar diffusivity $\kappa$ or the magnitude of the scalar gradient, where our analysis indicates the importance of the intrinsic length scale $\ell = (\nu \kappa / N^2 )^{1/4}$, which depends on fluid viscosity, scalar diffusivity, and scalar gradient.

The model predicts a maximum efficiency for ensembles of microorganisms which have no preferential swimming direction to be around $8\%$, which is achieved if $a/\ell=O(1)$. The efficiency $0.08$ might serve as an upper bound for the potential energy created in the ocean by microorganisms.  For example, using an approximate average Oxygen Utilization Rate (OUR) and the physics of respiration, \cite{Dewar:2006p11606} estimate that the total metabolic rate of all bacteria in the deep, unlit seas is about $6 \units{TW}$.  If we assume that all of these bacteria expend all of their energy in swimming and mix the ocean at maximum efficiency, this yields an upper bound for the creation of potential energy of $0.48 \units{TW}$.  

However, when calculated for realistic ocean parameters, we find that the value of $\ell$  is often large compared to the typical size of a bacterium and therefore the upper bound above is far too generous.  If $a/\ell\ll 1$ we  then find that the mixing efficiency is very small and even if a large portion of the bacterial metabolism is used in self-propulsion, the majority of this energy would be dissipated by viscous stresses.  For example, we might make a conservative estimate by assuming that all oceanic microorganisms   swim vertically (for example, gyrotactic algae)  and devote the majority of their metabolism to swimming.  If we suppose further that the ocean is strongly salt stratified with $N = 2 \units{cycles/hour} \approx 3.5 \times 10^{-3} \units{radians/sec}$, $\kappa \approx 10^{-9} \units{m^2/s}$, and $\nu = 1.6 \times 10^{-6} \units{m^2/sec}$, then $\ell = \left ( \nu \kappa / N^2 \right )^{1/4} \approx   3.4 \units{mm}$.  For a $10 \units{\mu m}$ size-organism we then obtain 
\begin{equation}\label{small}
\eta \approx 1.21 \left ( \frac{a}{\ell} \right )^3 \approx 3.1 \times 10^{-8},
\end{equation}
showing that the total contribution of microorganisms to ocean mixing is negligible.  

One possible objection to this estimate is that while $N = 2 \units{cycles/hour}$ is a high value for the average  overall stratification in the ocean measured over length scales of hundreds of meters, there exists significant microstructure and small-scale variation in stratification due to turbulent and disordered motion of fluid which may increase local gradients in density on the scales experienced by microorganisms.  In order to take this small-scale variation into account we can use estimates for the Cox number, $\Cx$, which is a non-dimensional measure of the small-scale variation of a scalar \citep{Gregg:1977p11639}. Using the example of  temperature,  $T$, we have 
\begin{equation}
\Cx= \frac{ \left \langle \left ( \partial T / \partial z  \right )^2 \right \rangle }{ \left \langle \partial T / \partial z \right \rangle^2},
\end{equation}
where the bracket $\langle \cdot \rangle$ denotes an average over some region in space.  The average variation in temperature gradient can be easily related to the average variation in mixing efficiency when $a / \ell \ll 1$ as
\begin{equation}
\langle \eta \rangle \approx 1.21 a^3 \left \langle \frac{1}{\ell^3} \right \rangle = 1.21 \frac{a^3}{\left ( \nu \kappa \right )^{3/4}} \left \langle N^{3/2} \right \rangle.
\end{equation}
for a microswimmer  oriented vertically.  In waters which are stratified by temperature the buoyancy frequency $N$ is proportional to $\left ( \partial T / \partial z \right )^{1/2}$, we can estimate $\llangle N^{3/2} \rrangle$ with the Cox number by assuming that $\llangle \left ( \partial T / \partial z \right )^{3/4} \rrangle \approx \llangle \left ( \partial T / \partial z \right )^2 \rrangle^{3/8}$.  This assumption will  tend to overestimate the effects of the small-scale variation of $\partial T / \partial z$ on $\llangle N^{3/2} \rrangle$ and thus overestimate the average mixing efficiency.  We then find that
\begin{equation}
\llangle N^{3/2} \rrangle \approx \left ( \Cx \, N_0^4 \right )^{3/8},
\end{equation}
where $N_0$ is the buoyancy frequency measured on large scales, which implies that the actual average mixing efficiency taking into account small-scale variations in density gradient might be estimated with
\begin{equation}
\llangle \eta \rrangle \approx \Cx^{3/8} \eta_0,
\end{equation}
where $\eta_0$ is the mixing efficiency estimated using the large-scale buoyancy frequency $N_0$.  A large Cox number would then naturally lead to an increase of the mixing efficiency from the value in \eqref{small}.  The highest Cox number measured by \cite{Gregg:1977p11639} over three cruises was $\Cx \approx 240$, leading to an increase by a factor of 8 above the result in \eqref{small}. One  would have to find an enhancement of mixing efficiency by at least five orders of magnitude for mixing by microorganisms to be geophysically relevant.  It seems therefore that even taking into account microstructure variations in density, the mixing efficiency of microswimmers is negligible.

An important point is that this conclusion does not prelude the possibility that larger organisms, which may produce fluid motions which possess larger $\Re$, larger $\Pe$, and larger $a / \ell$, are associated with appreciable mixing efficiencies.  This important question in biogenic mixing remains open.  The conclusion that bacteria do not contribute to ocean mixing confirms that \cite{Dewar:2006p11606} were correct to exclude bacteria from their assessment of the total contribution of swimming organisms to the mechanical energy budget of the ocean.

\section*{Acknowledgments}

This work was funded in part by the US National Science Foundation under OCE10-57838.

\appendix

\section{Efficiency scaling in the limit  $a \gg  \ell $}
\label{appendixScalings}

\subsection{Vertical Orientation}

When the distributed force $\bbf$ is oriented vertically and its size is increasing, we find a dominant balance in the momentum equation as $\rho_0 b \bzh \sim \bbf $ because the buoyancy term is increasing with $(a / \ell)^4$.  This means that $P_g = \rho_0 \int w b \d V \sim \int w f_3 \d V$.  With $\bbf \sim D / a^4$,  the momentum balance implies $b \sim D / \rho_0 a^4$ and therefore from buoyancy conservation $w \sim D \kappa / \rho_0 N^2 a^6$, which gives 
\begin{equation}
P_{\tot} \sim P_g = \rho_0 \int_{\R^3} w b \d V \sim \rho_0 \left ( \frac{D \kappa}{N^2 \rho_0 a^6} \right ) \left ( \frac{D}{\rho_0 a^4} \right ) a^3 = \frac{D^2 \ell^4}{\mu a^7},
\end{equation}
where we have substituted $\kappa / N^2 = \ell^4 / \nu$. Consequently, we have 
and $\eta = P_g / P_{\tot} \sim 1$, as seen computationally in  Fig.~\ref{mixingEfficiency}.

\subsection{Horizontal orientation}

In this case it is more difficult to find the gravitational potential energy because neither the buoyancy or vertical velocity is directly balanced by the distributed force.  For $u$ and $v$ we have the same balance $u \sim v \sim D / \rho_0 \nu a^2$.  To obtain  a relationship between   buoyancy and the velocity   we  take partial derivatives in $z$ and $y$ in the $y$-- and $z$--momentum equations respectively, subtract them, and then substitute for $w$ using the partial derivative in $x$ of the buoyancy conservation equation, which yields
\begin{equation}
\left ( \ell^4 \nabla^2 + 1 \right ) \frac{\partial b}{\partial y} = \nu \nabla^2 \left ( \frac{\partial u}{\partial z} \right ) \implies b \sim \frac{D}{\rho_0 a^4}\cdot
\end{equation}
We then find $w \sim D \ell^4 / \mu a^6$ from buoyancy conservation, which implies
\begin{equation}
P_g = \rho_0 \int_{\R^3} w b \d V \sim \rho_0 \left ( \frac{D \ell^4}{\mu a^6} \right ) \left ( \frac{D}{\rho_0 a^4} \right ) a^3 \sim \frac{D^2 \ell^4}{\mu a^7},
\end{equation}
and
\begin{equation}
P_{\tot} =  \int_{\R^3} u \left ( \bbf \cdot \bxh \right ) \d V \sim \left ( \frac{D}{\mu a^2} \right ) \left ( \frac{D}{a^4} \right ) a^3 \sim \frac{D^2}{\mu a^3}, 
\end{equation}
and therefore the efficiency scales as  \begin{equation}
\eta \sim \left ( \frac{a}{ \ell} \right )^{-4}\cdot
\end{equation}

\section{Fourier space integrals}
\label{appendixIntegrals}

\subsection{Vertically-oriented regularized dipole}

For the rate-of-creation of gravitational potential energy by a vertically-oriented regularized dipole of the form \eqref{F2}, for which $\bbeta = | \bbeta | \bzh$, we have
\begin{eqnarray}
P_g &=& \rho_0 \int_{\R^3} w^{\rdip}(\bx) b^{\rdip}(\bx) \d V \, ,  \nonumber \\
\nonumber &=& \frac{\rho_0}{8 \pi^3} \int_{\ksp} \! \! \! \! \! \! \! \! \! \tilde w^{\rdip}(\bk) \tilde b^{\rdip}(-\bk) \d \bk \, , \\
\nonumber &=& \frac{\rho_0}{8 \pi^3} \int_{\ksp} \! \! \! \! \! \! \! \! \! k_3^2 \tilde w^{\rstrat}(\bk) \tilde b^{\rstrat}(-\bk) \d \bk \, , \\
&=& \frac{D^2}{4 \pi^2 \nu \ell^3} \int_0^{\infty} V(k) e^{- (k \e)^2} \d k \, ,
\end{eqnarray}
where $k = | \bk |$ and $\e = a / \ell$, and $V(k)$ is defined in \eqref{vk}.  In the last step we have inserted the solutions, non-dimensionalized the integral by substituting $k = k' / \ell$ (and then dropping the primes for simplicity), converted to spherical coordinates, and integrated from 0 to $2 \pi$ over the azimuthal angle (in the vertical case, this integral is trivial as the solution is axisymmetric).  For the total rate-of-working on the fluid we find, in similar fashion,
\begin{eqnarray}
\nonumber P_{\tot} &=& \frac{1}{8 \pi^3} \int_{\ksp} \! \! \! \! \! \! \! \! \! k_3^2 \left [ \tilde \bu(\bk)^{\rstrat} \cdot \tilde \bbf(-\bk)^{\rstrat} \right ] \d \bk \, , \\
&=& \frac{D^2}{4 \pi^2 \mu \ell^3} \int_0^{\infty} W(k) e^{- (k \e)^2} \d k,
\end{eqnarray}
where $W(k)$ is defined in \eqref{wk}.

\subsection{Horizontally-oriented regularized dipole}

The rate-of-creation of gravitational potential energy by a vertically-oriented regularized dipole of the form \eqref{F2}, for which $\bbeta = | \bbeta  | \bxh$, following the calculation for the vertically-oriented regularized dipole, becomes
\beq
P_g = \frac{3 D^2}{32 \pi^2 \mu \ell^3} \int_0^{\infty} V(k) e^{- (k \e)^2} \d k \, ,
\label{horz_Pg}
\eeq
while the total rate-of-working on the fluid is given by
\begin{eqnarray}
\nonumber P_{\tot} &=& \frac{D^2}{4 \pi^2 \mu \ell^3} \Bigg [ \frac{1}{8} \int_0^{\infty} U(k) e^{- (k \e)^2} \d k  + \frac{1}{2} \int_0^{\infty} W(k) e^{- (k \e)^2} \d k \Bigg ],
\label{horz_Ptot}
\end{eqnarray}
where $U(k)$ is defined
\beq
\begin{split}
U(k) &= k^2 \left ( 1 + k^4 \right ) \int_0^{\pi} \frac{\sin^5 \psi}{k^4 + \sin^2 \psi} \d \psi, \\
&= \frac{2}{3} k^2 \left ( 2 - k^4 - 3 k^8 \right ) + 2 k^{10} \sqrt{1+k^4} \log \left [ \frac{1}{k^2} \left ( \sqrt{1+k^4} - 1 \right ) \right ].
\label{uk}
\end{split}
\eeq
Here the integrals are written after integration over the azimuthal angle.  The solution is no longer axisymmetric but this integral is still straightforward (and, instead of multiplying the expression by $ 2 \pi $ as in the axisymmetric case, produces other factors). 

\section{Ensemble of microswimmers with uniformly distributed orientations}
\label{appendixEnsemble}

To find the average of an ensemble of microswimmers with uniformly distributed orientations, we must first find the flow field for a regularized dipole with arbitrary orientation.  Consider the Stratlet oriented at an arbitrary angle $\alpha$ with respect to the vertical direction.  Because the governing equations are linear, we can express it  as a superposition of a horizontally-oriented Stratlet, $^H\bu^{\rstrat}$, and a vertically-oriented Stratlet, ${^V\bu^{\rstrat}}$,
\beq
\tilde \bu^{\rstrat} (\alpha) = {^H \tilde \bu^{\rstrat}} \sin \alpha + {^V\tilde \bu^{\rstrat}} \cos \alpha.
\eeq
For a dipole oriented at the same angle $\alpha$ we have that $\tilde \bu^{\rdip} = - i \left ( \bbeta \cdot \bk \right ) \tilde \bu^{\rstrat}$ with $\bbeta = D \left ( \sin \alpha \bxh + \cos \alpha \bzh \right )$ and where $\bxh$ and $\bzh$ are unit vectors in the horizontal and vertical directions, respectively.  The velocity field for the regularized dipole is then given by
\beq
\tilde \bu^{\rdip} \left ( \alpha \right ) = - i D \left ( k_1 \sin \alpha + k_3 \cos \alpha \right ) \left ( {^H \tilde \bu^{\rstrat}} \sin \alpha + {^V\tilde \bu^{\rstrat}} \cos \alpha \right ).
\eeq 
Expression for the other  variables such as the force distribution, $\tilde \bbf$, and the buoyancy field, $\tilde b$, follow similarly.  From these,  we are then able to calculate the rate-of-creation of gravitational potential energy and the total rate-of-working on the fluid using their definitions in \eqref{powerIntegralDefinitions}.  

For the rate-of-creation of gravitational potential energy we find
\begin{eqnarray}
\nonumber P_g &=& \frac{\rho_0}{8 \pi^3} \int_{\ksp} \tilde w(\bk) \tilde b(-\bk) \d \bk = - \frac{\rho_0 N^2}{8 \pi^3 \kappa} \int_{\ksp} \frac{1}{k^2} \tilde w (\bk) \tilde w (-\bk) \d \bk, \\
\nonumber &=& \frac{\rho_0 N^2 D^2}{8 \pi^3 \kappa} \int_{\ksp} \frac{1}{k^2} \Big [ k_1 {^H} \tilde w^{\rstrat}(\bk) \sin^2 \alpha + k_3 {^V} \tilde w^{\rstrat}(\bk) \cos^2 \alpha \\
&& \qquad  \qquad \qquad \qquad + \sin \alpha \cos \alpha \left ( k_1 {^V} \tilde w^{\rstrat}(\bk) + k_3 {^H} \tilde w^{\rstrat}(\bk) \right ) \Big ]^2 \d \bk,
\end{eqnarray}
where we have used the relation between $w$ and $b$ in \eqref{lowRePe_buoyancy} to write the integral solely in terms of the regularized Stokeslet velocity field.  Similarly, for the total rate-of-working on the fluid we obtain
\begin{eqnarray}
\nonumber P_{\tot} &=& \frac{D^2}{8 \pi^3} \int_{\ksp} {^H} \tilde u_1^S \left ( k_1^2 \sin^4 \alpha + k_3^2 \sin^2 \alpha \cos^2 \alpha \right ) + 2 \left ( {^V} \tilde u_1^S + {^H} \tilde u_3^S \right ) k_1 k_3 \sin^2 \alpha \cos^2 \alpha \\
\nonumber&&   \qquad\qquad\qquad + {^V} \tilde u_3^S \left ( k_1^2 \sin^2 \alpha \cos^2 \alpha + k_3^2 \cos^4 \alpha \right ) \d \bk  \\
\nonumber&=& {^H} P_{\tot} \sin^4 \alpha + {^V} P_{\tot} \cos^4 \alpha \\
&&    + \sin^2 \alpha \cos^2 \alpha \left [ \frac{D^2}{8 \pi^3} \int_{\ksp} k_3^2 {^H} \tilde u^{\rstrat} + 2 k_1 k_3 \left ( {^V} \tilde u^{\rstrat} + {^H} \tilde w^{\rstrat} \right ) + k_1^2 {^V} \tilde w^{\rstrat} \d \bk \right ].
\end{eqnarray}

An ensemble average, $\Pavg$, can be found by integrating each energetic rate over all orientations following
\begin{equation}
\Pavg = \frac{1}{4 \pi} \int_0^{2 \pi} \int_0^{\pi} P(\alpha) \sin \alpha \d \alpha \d \theta = \frac{1}{2} \int_0^{\pi} P(\alpha) \sin \alpha \d \alpha.
\end{equation}
For the ensemble-averaged rate of creation of gravitational potential energy  we then find
\beq
\begin{split}
\Pavg_{g} &= \frac{8}{15} {^H} P_{g} + \frac{1}{5} {^V} P_{g} \\
& \qquad + \frac{2}{15} \left [ \frac{\rho_0 N^2 D^2}{8 \pi^3 \kappa} \int_{\ksp} \frac{1}{k^2} \left ( k_1^2 \left ( {^V} \tilde w^{\rstrat} \right )^2 + 4 k_1 k_3 \left ( {^V} \tilde w^{\rstrat} \right ) \left ( {^H} \tilde w^{\rstrat} \right ) + k_3^2 \left ( {^H} \tilde w^{\rstrat} \right )^2 \right ) \d \bk \right ],
\end{split}
\eeq
and for the ensemble-averaged total rate-of-working on the fluid we have
\begin{equation}
\Pavg_{\tot} = \frac{8}{15} {^H} P_{\tot} + \frac{1}{5} {^V} P_{\tot} + \frac{2}{15} \left [ \frac{D^2}{8 \pi^3} \int_{\ksp} k_3^2 {^H} \tilde u^{\rstrat} + 2 k_1 k_3 \left ( {^V} \tilde u^{\rstrat} + {^H} \tilde w^{\rstrat} \right ) + k_1^2 {^V} \tilde w^{\rstrat} \d \bk \right ],
\end{equation}
all of which can easily be evaluated computationally from the analytical solutions. 

\section{Asymptotic evaluation of mixing efficiency}
\label{appendixAsymptotics}

It it possible to mathematically analyze the integrals for $P_{\tot}$ and $P_g$ asymptotically in the limits $a / \ell \ll 1$ and $a / \ell \gg 1$ in order to generate  approximate formulae for the mixing efficiency.  Here we examine the integrals for the vertically-oriented regularized force dipole as an example; the approach for the asymptotic evaluation of the horizontal integrals  is similar.  We use the shorthand $\e = a / \ell$ to simplify the expressions.

The total rate-of-work, $P_{\tot}$, and the rate-of-creation of gravitational potential energy, $P_g$, for the 
vertically-oriented regularized force dipole are given by
\begin{equation}
P_g = \frac{D^2}{4 \pi^2 \mu \ell} \int_0^{\infty} V(k) e^{- (\e k)^2} \d k,
\end{equation}
and
\begin{equation}
P_{\tot} = \frac{D^2}{4 \pi^2 \mu \ell} \int_0^{\infty} W(k) e^{- (\e k)^2} \d k,
\end{equation}
where we have defined $\e = a / \ell$ and $V(k)$ and $W(k)$ are defined by Equations \ref{vk} and \ref{wk}, respectively.  The coefficients in front of the integrals are identical so we need only consider the ratio of the integrals.

\subsection{Small-organism and weak stratification limit, $a / \ell \ll 1$}

We find the first two terms  in an asymptotic expansion of the integral for $\e = a / \ell \ll 1$.  Because $W(k)$ is divergent, the leading order contribution as $\e \ll 1$ will come from large values of $k$.  As such we attempt to ``divide and conquer'' the integrals by splitting them into a local contribution for large $k$ and a global contribution for the rest.  For the integral for the rate of creation of gravitational potential energy $I_g$ we write
\beq
I_g = \int_0^{\infty} V(k) e^{- (\e k)^2} \, \d k = \int_0^M V(k) e^{- (\e k)^2} \d k + \frac{1}{\e} \int_{\e M}^{\infty} V \left ( u / \e \right ) e^{-u^2} \d u = I_{g,G} + I_{g, L},
\eeq
where we take $M \gg 1$ as $\e \ll 1$.  We find for the local contribution
\beq
\begin{split}
I_{g,L} &= \frac{1}{\e} \int_{M \e}^{\infty} \left ( \frac{16}{105} \frac{\e^2}{u^2} - \frac{64}{315} \frac{\e^6}{u^6} + O(\e^{10} / u^{10}) \right ) e^{-u^2} \d u, \\
&= \frac{16}{105} \frac{1}{M} - \frac{64}{1575} \frac{1}{M^5} - \frac{16 \sqrt{\pi}}{105} \e + \frac{64}{945} \frac{\e^2}{M^3} - \frac{16}{105} M \e^2 + O(M^3 \e^4, \e^4/M).
\end{split}
\eeq
Because $V(k)$ vanishes as $k \gg 1$, we surmise that the global contribution is a constant at leading order.  We calculate this contribution numerically.  Because $V(k)$ involves the cancellation of very large terms at large $k$, the numerical integration is aided by {patching} the integral to its Taylor series expansion around $k \gg 1$.  We find
\begin{equation}
\int_0^M V(k) \d k + \frac{16}{105} \frac{1}{M} - \frac{64}{1575} \frac{1}{M^5} + O \left (\frac{1}{M^9} \right ) = \int_0^{\infty} V(k) \d k \simeq 0.143313,
\end{equation}
and therefore
\beq
I_g \left ( \e \ll 1 \right ) = 0.1433 - \frac{16 \sqrt{\pi}}{105} \e + O \left ( \e^2 \right ).
\eeq
The integral for the total rate-of-working $I_{\tot}$ is approached in the same manner.  For the local integral we find
\beq
I_{\tot,L} = \frac{1}{\e} \int_{M \e}^{\infty} \left ( \frac{4}{15} \frac{u^2}{\e^2} - \frac{16}{105} \frac{\e^2}{u^2} \right ) e^{-u^2} \d u = \frac{\sqrt{\pi}}{15} \frac{1}{\e^3} - \frac{4}{45} M^3 + \frac{4}{75} M^5 \e^2 + O(M^7 \e^4).
\eeq
In the global integral we surmise that the largest term will cancel the $4 M^3 / 45$ term in the local integral and that the next order term will contribute a constant.  We find this constant by subtracting $4 k^2 / 15$ from the integrand, which cancels the largest contribution to $W(k)$, and numerically evaluating the remaining integral using the same  method used for $I_g$.  We find 
\beq
\int_0^{\infty} \left ( W(k) - \frac{4}{15} k^2 \right ) \d k = \int_0^M \left ( W(k) - \frac{4}{15} k^2 \right ) \d k - \frac{16}{105} \frac{1}{M} + \frac{32}{1575} \frac{1}{M^5} + O \left ( \frac{1}{M^9} \right ) = -0.191089.
\eeq
We therefore have 
\beq
I_{\tot} \left ( \e \ll 1 \right ) = \frac{\sqrt{\pi}}{15} \frac{1}{\e^3} - 0.1911 + O(\e).
\eeq
The efficiency $\eta$ is therefore given by
\beq
\eta \left ( \e \ll 1 \right ) = \frac{I_g \left ( \e \ll 1 \right )}{I_{\tot} \left ( \e \ll 1 \right )} = 1.212 \e^3 - 2.286 \e^4 + O \left ( \e^5 \right ),
\eeq
which agrees well with the full calculation of mixing efficiency. The calculation for the horizontal integrals is slightly more involved because there are two terms in the expression for the total rate of work, but the approach is identical, and we find
\beq
\eta_{\text{horz}} \left ( \e \ll 1 \right ) = 0.1516 \e^3 - 0.2857 \e^4 + O \left ( \e^5 \right ).
\eeq

\subsection{Large-organism and strong stratification limit, $a / \ell \gg 1$}
The limit $a / \ell \gg 1$ is easier than $a / \ell \ll 1$.  Again we use the shorthand $\e = a / \ell$.  As $\e \gg 1$, the integrand will be very small except in a small region around $k = 0$.  Thus all that is required is to expand the integrands around $k = 0$ and integrate term by term.  For $I_g$ we find
\beq
\begin{split}
I_g(\e \gg 1) &= \int_0^{\infty} V(k) e^{- (k \e)^2} \d k, \\
&= \int_0^{\infty} \left ( \frac{2}{3} k^6 + 8 k^{10} \log (k) + \left ( 5 - 2 \log (4) \right ) k^{10} + O(k^{14},k^{14} \log(k)) \right ) e^{- (k \e)^2} \d k, \\
&= \frac{5 \sqrt{\pi}}{8} \e^{-7} - \frac{945 \sqrt{\pi}}{8} \e^{-11} \log (\e) + 226.953 \e^{-11} + O(\e^{-15},\e^{-15} \log (\e)).
\end{split}
\eeq
For $I_{\tot}$ we find
\beq
\begin{split}
I_{\tot}(\e \gg 1) &= \int_0^{\infty} W(k) e^{- (k \e)^2} \d k, \\
&= \int_0^{\infty} \left ( \frac{2}{3} k^6 + 4 k^{10} \log(k) + \left ( 2 - \log(4) \right ) k^{10} + O(k^{14},k^{14} \log{k}) \right ) e^{- (k \e)^2} \d k, \\
&= \frac{5 \sqrt{\pi}}{8} \e^{-7} - \frac{945 \sqrt{\pi}}{16} \e^{-11} \log (\e) + 100.391 \e^{-11} + O(\e^{-15},\e^{-15} \log (\e)).
\end{split}
\eeq
The efficiency $\eta$ is therefore
\beq
\eta \left ( \e \gg 1 \right ) = \frac{ I_g \left ( \e \gg 1 \right )}{I_{\tot} \left ( \e \gg 1 \right )} = 1 - \frac{945}{10} \e^{-4} \log \left [ \e \right ] + O \left ( \e^{-8} \right ).
\eeq
The calculation for the horizontally-oriented swimmer can be approached in the same way, which yields
\beq
\eta_{\text{horz}} \left ( \e \gg 1 \right ) = \frac{15}{8} \e^{-4} + \frac{2835}{8} \e^{-8} \log \left [ \e \right ] + O  \left ( \e^{-8} \right ) .
\eeq

\subsection{Summary of asymptotic calculations}

As $\e = a / \ell \ll 1$, we find that the mixing efficiency of a vertically-oriented swimmer is
\beq
\eta_{\text{vert}} \left ( \frac{a}{\ell} \ll 1 \right ) = 1.212 \left ( \frac{a}{\ell} \right )^3 - 2.286 \left ( \frac{a}{\ell} \right )^4 + O \left ( \frac{a}{\ell} \right )^5.
\eeq
For the mixing efficiency of a horizontally-oriented swimmer as $\e = a / \ell \ll 1$ we find
\beq
\eta_{\text{horz}} \left ( \frac{a}{\ell} \ll 1 \right ) = 0.1516 \left ( \frac{a}{\ell} \right )^3 - 0.2857 \left ( \frac{a}{\ell} \right )^4 + O \left ( \frac{a}{\ell} \right )^5.
\eeq
As $\e = a / \ell \gg 1$, we find for the vertically-oriented swimmer
\beq
\eta_{\text{vert}} \left ( \frac{a}{\ell} \gg 1 \right ) = 1 - \frac{945}{10} \left ( \frac{1}{a / \ell} \right )^4 \log \left [ \frac{a}{\ell} \right ] + O \left (  \frac{1}{a / \ell} \right )^8 .
\eeq
For the horizontally swimmer as $\e = a / \ell \gg 1$,
\beq
\eta_{\text{horz}} \left ( \frac{a}{\ell} \gg 1 \right ) = \frac{15}{8} \left ( \frac{1}{a / \ell} \right )^4 + \frac{2835}{8} \left ( \frac{1}{a / \ell} \right )^8 \log \left [ \frac{a}{\ell} \right ] + O  \left ( \frac{1}{a / \ell} \right )^8 .
\eeq
In all four cases, the asymptotic results agree quantitatively with our numerical computations.

\bibliographystyle{JMR}
\bibliography{refs,refs2,books}

\end{document}